\documentclass[a4paper,11pt]{article}
\usepackage{subfig}
\usepackage{jcappub} 

\usepackage[T1]{fontenc} 

\title{Mass radius relation of compact stars in the braneworld}

\author[a,b,1]{Luis B. Castro,\note{Corresponding author.}}
\author[c]{Marcelo D. Alloy,}
\author[a,d]{D\'ebora P. Menezes
}

\affiliation[a]{Depto de F\'isica, CFM, Universidade Federal de Santa Catarina,\\Florian\'opolis, SC CP-476, CEP 88.040-900, Brazil}
\affiliation[b]{Departamento de F\'{\i}sica, Universidade Federal do Maranh\~{a}o, \\Campus Universit\'{a}rio do Bacanga, CEP 65085-580, S\~{a}o Lu\'{i}s, Maranh\~{a}o, Brazil}
\affiliation[c]{Universidade Federal da Fronteira Sul,\\Chapec\'o, SC, CEP 89.812-000, Brazil}
\affiliation[d]{Departamento de F\'isica Aplicada, Universidad de Alicante,\\Ap. Correus 99, E-03080, Alicante, Spain}

\emailAdd{luis.castro@pgfsc.ufsc.br}
\emailAdd{alloy@uffs.edu.br}
\emailAdd{debora.p.m@ufsc.br}

\abstract{The braneworld scenario, based on the fact that the four
  dimension space-time is a hyper-surface of a five dimensional
  manifold, was shown to deal in a satisfactory way with the hierarchy
  problem. In this work we study macroscopic stellar properties of
  compact stars from the braneworld point of view. Using neutron star
  equations of state, we test the possibility of extra dimensions by
  solving the brane Tolman-Oppenheimer-Volkoff equations obtained for
  three kinds of possible compact objects: hadronic, hybrid and quark
  stars. By comparing the macroscopic solutions with observational
  constraints, we establish a brane tension lower limit and the value for which the
  Tolman-Oppenheimer-Volkoff equations in the braneworld converge to
  the usual Tolman-Oppenheimer-Volkoff equations.}

\arxivnumber{1403.1099}

\dedicated{Dedicated to my family, Isabel and Alexander\\ (L. B. Castro)}

\begin{document}
\maketitle
\flushbottom

\section{Introduction}
\label{sec:intro}

The four fundamental nature interactions (gravity, electromagnetism,
strong and weak nuclear forces) were investigated separately
during their initial studies and developments. At some point,
it became obvious that an unified theory could help the understanding
of the primordial universe and its evolution. Two of these fundamental
forces (electromagnetism and weak nuclear forces) were then joined in
an unique formalism known as the electroweak interaction by Abdus Salam,
Sheldon Glashow and Steven Weinberg, but the unification of the other
two is still pursued nowadays. Around 1990's, string theory came up
as a possible candidate for the {\it theory of everything} or
{\it M-theory} because it presented
the advantage of conciliating quantum mechanics with gravity by putting aside
the idea of particles as the elementary bricks of the universe and explaining
them as particular quantum states of strings. String theory started with
bosons only and later incorporated fermions and the requirement of extra
dimensions was a key point in its development. M-theory assumes the existence
of 11 dimensions, some of them treated as hidden dimensions.
These extra dimensions of the space-time have since then been used as an attempt
to explain several open problems like the existence of dark matter,
dark energy, the accelerated expansion of the universe, etc.

The braneworld scenario is based on the assumption that the
four dimensional space-time is a hyper-surface of a five dimensional manifold.
The gravity in
the braneworld is a five dimensional phenomenon and the other fundamental
interactions are confined to the brane. The braneworld scenario has attracted
a lot of interest, because
it tackles the hierarchy problem~\cite{rs1} in an effective way. Another
attractive property is that the Newtonian law of gravity with a correction
is also given in this braneworld scenario~\cite{rs2}. In the Randall-Sundrum
(RS) model~\cite{rs1}, one can further add scalar fields~\cite{wise} with usual dynamics and allow them to interact with gravity in the standard way. In this scenario, the smooth character of the solutions generate thick branes with a diversity of structures~\cite{townsend,tow1,tow2,tow3}. In the braneworld scenario an import issue is how gravity and different observable matter fields of the Standard Model of particle physics are localized on the brane. Recently, the localization of fermions and bosons on the brane have received considerable attention in the
literature~\cite{shapo,sha1,sha2,sha3,sha4,single,sin1,sin2,mel,chino1,chi1,chi2,casana,ben1,ben2,landim,lan1,lan2,cruz2,cruz2_1,cruz2_2}.

On the other hand, in~\cite{germani,bernhardt,ovalle1}, the idea that neutron stars
could be a laboratory
for testing extra dimensions was first explored. It was found that the star
in the braneworld is less compact that in general relativity and an
astrophysical lower limit for the brane tension was established. In~\cite{garcia}, brane theory was utilized to estimate the correction in
the waves radiated by a binary system and the observational masses of the
PSR B1913+16 were then used to constrain the brane tension.

Original neutron star models assume that the dense matter in its interior
is composed of hadrons (protons and neutrons only or the complete lowest
lying baryonic octet) and leptons, responsible for ensuring charge neutrality
and $\beta$-equilibrium~\cite{glen}. On the other hand, the Bodmer-Witten
conjecture~\cite{Bodmer_Witten_Itoh,Bodmer_Witten_Itoh2,Bodmer_Witten_Itoh3} states that quarks can be deconfined
from the hadrons, forming a stable quark matter under certain conditions.
Hence, compact stars can be constituted of pure quark matter or
perhaps of hybrid matter, containing in their core a pure quark phase or a
mixed phase of quarks and hadrons~\cite{glen,mp1,mp1_1,mp2,bielich,yang}.
In order to rule out improbable stellar configurations, observational
constraints have been used. While most neutron stars have masses of the order
of 1.4 M$_\odot$, at least two pulsars, PSR J1614-2230~\cite{demorest} and
PSR J0348+0432~\cite{antoniadis} were confirmed to be very massive objects,
with masses of the order of $2 M_\odot$. The theoretical calculation of
macroscopic stellar properties, as the masses and radii, is done by solving the
Tolman-Oppenheimer-Volkoff (TOV)~\cite{tov} equations, which use equations of
state (EOS) as input.

In this work, we study the effects of the braneworld scenario on the macroscopic stellar properties of compact stars. In that spirit, we revisit the idea of solving a TOV-like system of equations in the braneworld (brane-TOV) and see if the resulting mass and radius results survive the known observational constraints. We also check weather, with more realistic equations of state than the perfect fluid one used in~\cite{germani,garcia}, limits for the brane tension can be established. Furthermore, we show that the star becomes more compact and that the radii can be adjusted to smaller values depending of the brane-TOV parameters.

The paper is organized as follows: In Sec.~\ref{sec2}, we present the standard TOV formalism. In Sec.~\ref{sec3}, we give a brief review of the brane-TOV formalism. In Sec.~\ref{sec4}, we use different EOS as input to the brane-TOV equations, present and discuss our results. Finally, in Sec.~\ref{sec5}, we draw our final conclusions.

\section{Tolman-Oppenheimer-Volkoff Equations}
\label{sec2}

The Tolman-Oppenheimer-Volkoff (TOV) equations for static and
spherical stars are derived from the standard 4D general relativity~\cite{tov}
\begin{equation}
G_{\alpha \beta}=R_{\alpha \beta}-\frac{1}{2}Rg_{\alpha \beta}=\kappa^2 T_{\alpha \beta},
\end{equation}

\noindent where $\kappa^{2}=8\pi G$, $g_{\alpha \beta}$ is the metric,
$R_{\alpha \beta}$ is the Ricci tensor,
$R=g^{\alpha \beta}R_{\alpha \beta}$ is the scalar curvature
and $T_{\alpha \beta}$ is the energy-momentum tensor. The differential element
for a spherical relativistic star is given by
\begin{equation}\label{difel}
ds^2=-e^{2\phi(r)}dt^2+e^{2\Lambda(r)}dr^2+r^2(d\theta^2+\sin^2\theta d\phi^2).
\end{equation}

\noindent Solving Einstein's field equations for a perfect fluid matter, we obtain
\begin{eqnarray}
\frac{dm}{dr}&=&4\pi r^2\epsilon,\\
\frac{dp}{dr}&=&-(\epsilon+p)\frac{d\phi}{dr},\\
\frac{d\phi}{dr}&=&\frac{Gm+4\pi Gr^3p}{r(r-2m)},\\
\nonumber
\end{eqnarray}

\noindent where $m(r)$ and $p(r)$ are respectively the gravitation mass and pressure
both defined as star radius functions and $\epsilon$ is the energy density. In this case, we have 3 equations and 4 unknown functions, which are $m(r)$, $\epsilon(r)$, $p(r)$ and $\phi(r)$. For this reason, to solve these equations, we need an equation of state $p(\epsilon)$
and appropriate boundary conditions, which we choose as:
\begin{equation}
m(0)=0,\qquad p(0)=p_c,\qquad m(R)=M,\qquad p(R)=0,
\end{equation}

\noindent where $p_c$ is central pressure, R is the star radius and M is the
total gravitation mass of the star.

\section{Tolman-Oppenheimer-Volkoff Equations in the braneworld}
\label{sec3}

According to the current idea, our observable universe may be confined on some hypersurface (4D) called brane which, in turn, is embedded in some multidimensional space (5D) called bulk. In this context, only gravitation could propagate in the bulk (extra-dimension). Within the braneworld scenario, the field equations induced on the brane are derived via an elegant geometric approach developed in Ref.~\cite{shiro}. The basic idea of this approach is to project the 5D curvature along the brane and the result is a modification of the standard Einstein's equations, with the new terms carrying bulk corrections onto the brane. The bulk corrections to Einstein's equations on the brane can be consolidated into an effective total energy-momentum tensor~\cite{shiro}
\begin{equation}
G_{\alpha \beta}=\kappa^2 T_{\alpha \beta}^{\mathrm{eff}},
\end{equation}

\noindent where $G_{\alpha\beta}$ is the usual Einstein's tensor and
\begin{equation}\label{teff}
    T_{\alpha\beta}^{\mathrm{eff}}=T_{\alpha\beta}+\frac{6}{\lambda}\,S_{\alpha\beta}-\frac{1}{\kappa^2}\,\mathcal{E}_{\alpha\beta}\,,
\end{equation}

\noindent where $\lambda$ is the brane tension, which correspond to the vacuum energy density on the brane. The effective total energy-momentum tensor~\eqref{teff} show two key modifications to the standard 4D Einstein's field equations. The bulk corrections can be classified as local and non-local corrections. The local correction is carried via the tensor $S_{\alpha\beta}$ (matter corrections), while the non-local correction is carried via the projection $\mathcal{E}_{\alpha\beta}$ of the bulk Weyl tensor. The geometric tensor $\mathcal{E}_{\alpha\beta}$ transmit non-local gravitational degrees of freedom from the bulk to the brane.

For a perfect fluid or minimally coupled scalar field the expressions for $T_{\alpha \beta}$ and $S_{\alpha \beta}$ are given by~\cite{maar,maar_2}
\begin{eqnarray}
  T_{\alpha\beta} &=& \rho u_{\alpha}u_{\beta}+ph_{\alpha\beta}, \\
  S_{\alpha\beta} &=& \frac{1}{12}\,\rho^{2}u_{\alpha}u_{\beta}+\frac{1}{12}\,
\rho(\rho+2p) h_{\alpha\beta},
\end{eqnarray}

\noindent where $u^{\alpha}$ is the four-velocity and $h_{\alpha\beta}=g_{\alpha\beta}+u_{\alpha}u_{\beta}$ is the projection orthogonal to $u^{\alpha}$. Further, assuming static spherical symmetry the expression for $\mathcal{E}_{\alpha\beta}$
becomes~\cite{germani}:
\begin{equation}\label{eweyl}
    \mathcal{E}_{\alpha\beta}=-\frac{6}{\kappa^{2}\lambda}\,\left[ \mathcal{U}u_{\alpha}u_{\beta}+
    \mathcal{P}r_{\alpha}r_{\beta}+\frac{(\mathcal{U}-\mathcal{P})}{3}\,h_{\alpha\beta} \right]\,,
\end{equation}

\noindent where $r_{\alpha}$ is a unit radial vector, $\mathcal{U}$ and $\mathcal{P}$ are respectively the non-local energy density and non-local pressure on the brane. The terms $\mathcal{U}$ and $\mathcal{P}$ may be really interpreted as an energy density and pressure respectively, the label of "non-local" in the Weyl terms are associated with the fact that they have an extra dimensional origin and are commonly referred in the literature as "dark radiation" $\mathcal{U}$ and "dark pressure" $\mathcal{P}$~\cite{maar_2,ovalle2}. From~\eqref{eweyl}, we see that $\mathcal{E}_{\alpha\beta}\rightarrow0$ as $\lambda^{-1}\rightarrow0$. Applying this limit in eq.~\eqref{teff}, we obtain that $T_{\alpha\beta}^{\mathrm{eff}}=T_{\alpha\beta}$, therefore the standard 4D general relativity is regained.

Solving Einstein's equations on the brane, we obtain the modified TOV equations
\begin{eqnarray}
\frac{dm}{dr}&=&4\pi r^2\epsilon_{\mathrm{eff}},\label{eq1aa}\\
\frac{dp}{dr}&=&-(\epsilon+p)\frac{d\phi}{dr},\label{eq2aa}\\
\frac{d\phi}{dr}&=&\frac{Gm+4\pi Gr^3\left(p_{\mathrm{eff}}+\frac{4\mathcal{P}}{\kappa^{4}\lambda}\right)}{r(r-2Gm)},\label{eq3aa}\\
\frac{d\mathcal{U}}{dr}+\left( 4\mathcal{U}+2\mathcal{P} \right)\frac{d\phi}{dr}&=&-2(4 \pi G)^2 (\epsilon+p)
\frac{d \epsilon}{dr}-2\frac{d\mathcal{P}}{dr}-\frac{6}{r}\mathcal{P}\,,\label{eq4aa}
\end{eqnarray}

\noindent where
\begin{eqnarray}
  \epsilon_{\mathrm{eff}} &=& \epsilon+\frac{\epsilon^2}{2\lambda}+\frac{6}{\kappa^{4} \lambda}\mathcal{U}\,, \\
  p_{\mathrm{eff}} &=& p+\frac{p\epsilon}{\lambda}+\frac{\epsilon^2}{2\lambda}+\frac{2}{\kappa^{4} \lambda}\mathcal{U} \,.
\end{eqnarray}

\noindent In contrast to standard 4D general relativity, in this case
we have 4 equations and 6 unknown functions, which are $m(r)$,
$\epsilon(r)$, $p(r)$, $\phi(r)$, $\mathcal{U}(r)$ and
$\mathcal{P}(r)$. In order to solve satisfactorily these system of
equations, we need an equation of state $p=p(\epsilon)$ and
additionally an equation of state-like relation
$\mathcal{P}=\mathcal{P}(\mathcal{U})$. We assume that the Weyl terms
(non-local energy density and non-local pressure on the brane) obey
the simplest relation $\mathcal{P}=w \mathcal{U}$, in such a way that the system of equations~\eqref{eq1aa},~\eqref{eq2aa},~\eqref{eq3aa} and~\eqref{eq4aa} become:
\begin{eqnarray}
\frac{dm}{dr}&=&4\pi r^2\epsilon_{\mathrm{eff}},\label{eq1}\\
\frac{dp}{dr}&=&-(\epsilon+p)\frac{d\phi}{dr},\label{eq2}\\
\frac{d\phi}{dr}&=&\frac{Gm+4\pi Gr^3\left(p_{\mathrm{eff}}+\frac{4w}{\kappa^{4}\lambda}\mathcal{U}\right)}{r(r-2Gm)},\label{eq3}\\
\frac{d\mathcal{U}}{dr}&=&-\frac{2}{1+2w}\left[(4 \pi G)^2 (\epsilon+p)
\frac{d \epsilon}{dr}+ A\right]\label{eq4}\,,
\end{eqnarray}

\noindent where $A=\frac{3w}{r}\mathcal{U}+(2+w)\mathcal{U} \frac{d\phi}{dr}$. Finally, we need appropriate boundary conditions, which we choose as
\begin{equation}
\label{eq:x}
\begin{split}
m(0) &= 0 \,,
\qquad
p(0) = p_c \,,
\qquad
p(R) = 0 \,,
\\
m(R) &= M \,,
\qquad
\mathcal{U}(0) = 0 \,.
\end{split}
\end{equation}

Other choices for the initial condition on $\mathcal{U}$ are
 possible, but the one used and written above is
 numerically convenient and it may be interpreted as a condition that
 does not influence the non-local energy density at the center of the
 compact star. It is worth mentioning that different values of the initial
  condition for $\cal U$ lead to the same
final results after just few iterations, as explained in
\cite{bernhardt}.\footnote{In this case, \emph{iterations} refers to radial integration steps of the modified TOV equations.} 
We have checked that this is numerically correct.

To solve the Tolman-Oppenheimer-Volkoff (TOV) equations numerically
in the braneworld we use mixed units \cite{glen}, which are useful in
stellar calculations, where pressure and energy density are given in
km$^{-2}$ units and $w$ is a dimensionless quantity. The unit of the brane tension $\lambda$ is also km$^{-2}$, but for convenience we present
our results in dyn/cm$^{2}$, the unit generally used in the literature (see~\cite{garcia}, for instance).

The solutions from equations \eqref{eq1}, \eqref{eq2}, \eqref{eq3} and
\eqref{eq4} can be separated into internal and external solutions. The
internal solution is the result obtained by integrating the system from the
core of the star until the null pressure point, at its surface.
The external solution is obtained from the null pressure until the Weyl
term $\mathcal{U}$ becomes approximately null. In the present work we
restrict ourselves to the study of the internal solutions.

\section{Testing the equations of state}
\label{sec4}

We next choose some EOS as input to the brane-TOV equations. As mentioned
in the Introduction, depending on their possible interior composition, neutron
stars can be classified as hadronic stars with or without hyperons~\cite{glen,luiz}, hybrid stars containing hadronic and quark phases~\cite{mp1,mp1_1,mp2,paoli} or hadronic and pion or kaon condensates~\cite{condensates,condensates_1,condensates_2,condensates_3}
and quark (also known as strange) stars~\cite{quark,james}.
In the following all three possibilities are analyzed. We start from the
case in which just the solution related to the interior of the star is
considered. For hadronic and hybrid stars, a crust is always expected and
hence, we add the Baym-Pethick-Sutherland (BPS)~\cite{bps} equation of state for very low densities.
Due to the recent discoveries of massive stars~\cite{demorest,antoniadis}
we start our analysis from EOS that we know are capable of generating high
maximum masses when used as input to the usual TOV equations.

\begin{figure}[tbp]
\centering
\subfloat[]{\includegraphics[width=0.45\linewidth,angle=0]{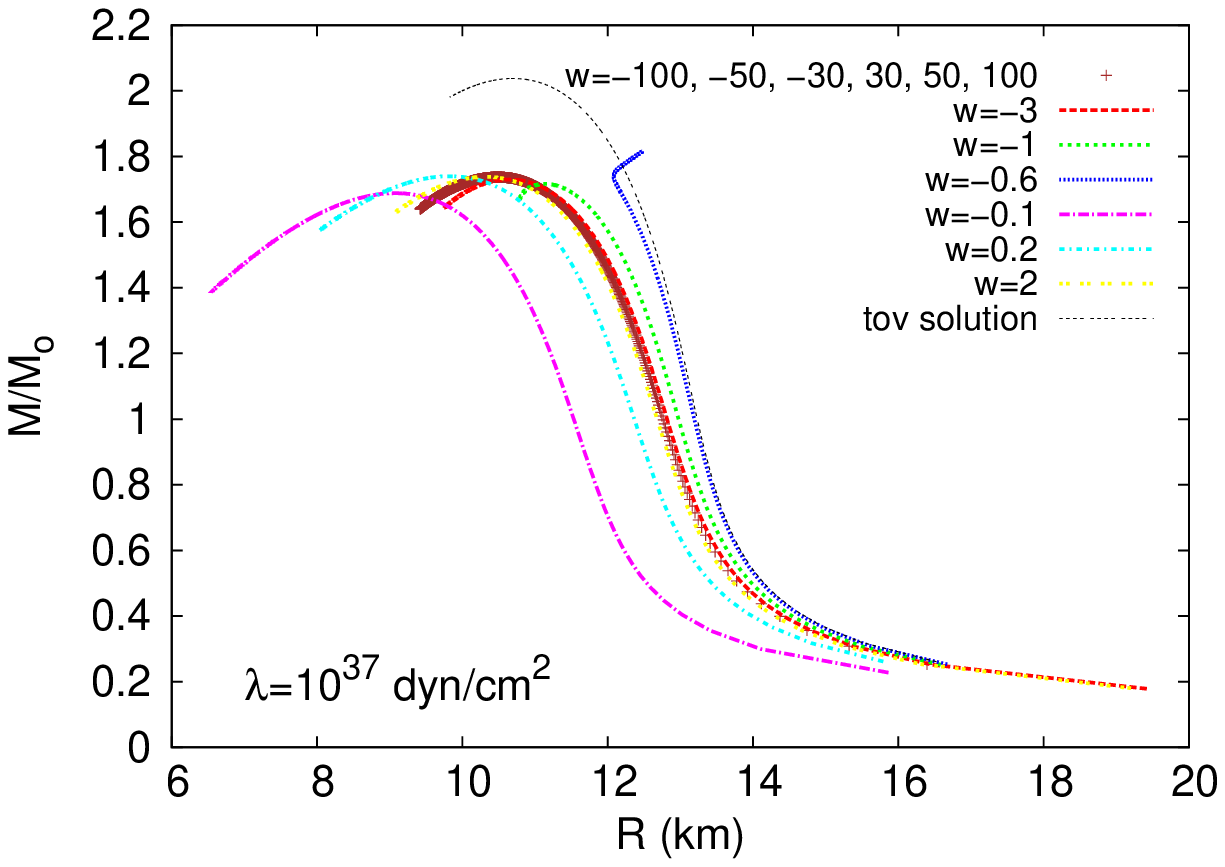}\label{fighad:a}}
\hfill
\subfloat[]{\includegraphics[width=0.45\linewidth,angle=0]{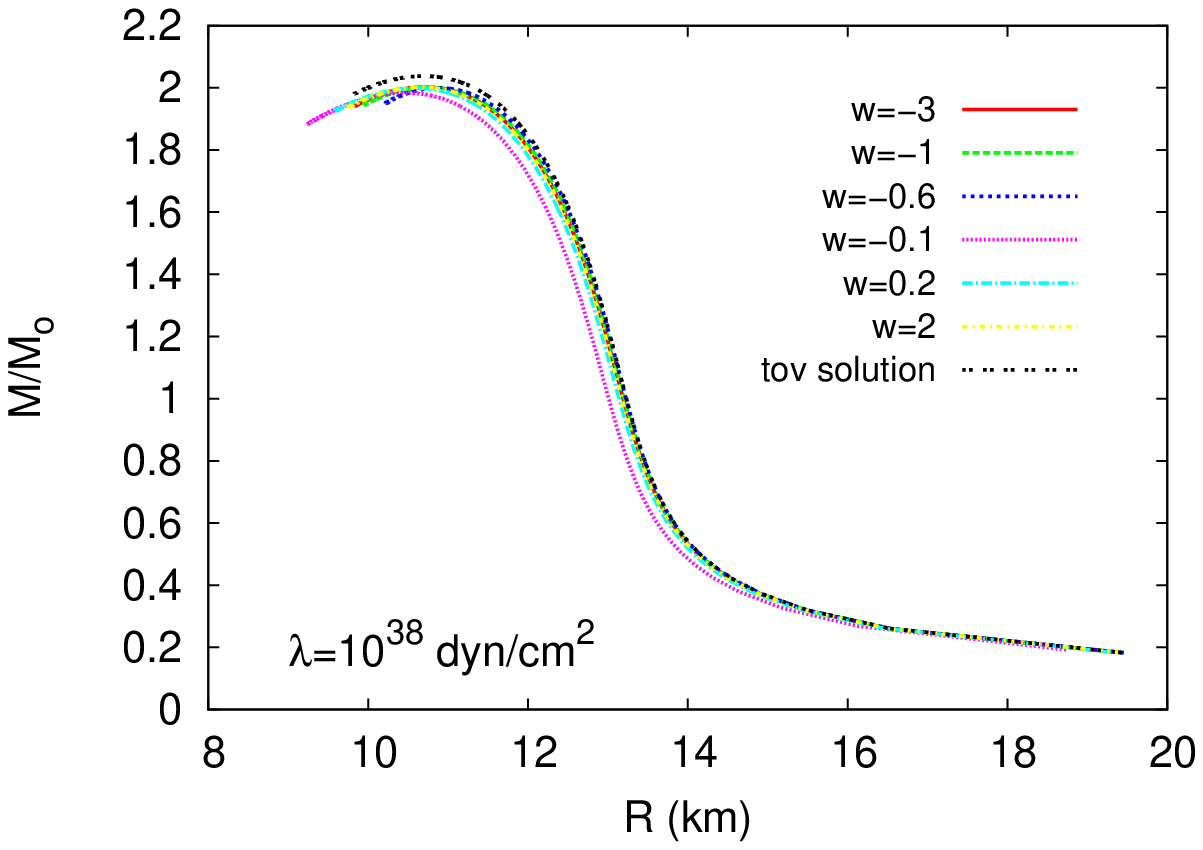}\label{fighad:b}}
\hfill
\subfloat[]{\includegraphics[width=0.45\linewidth,angle=0]{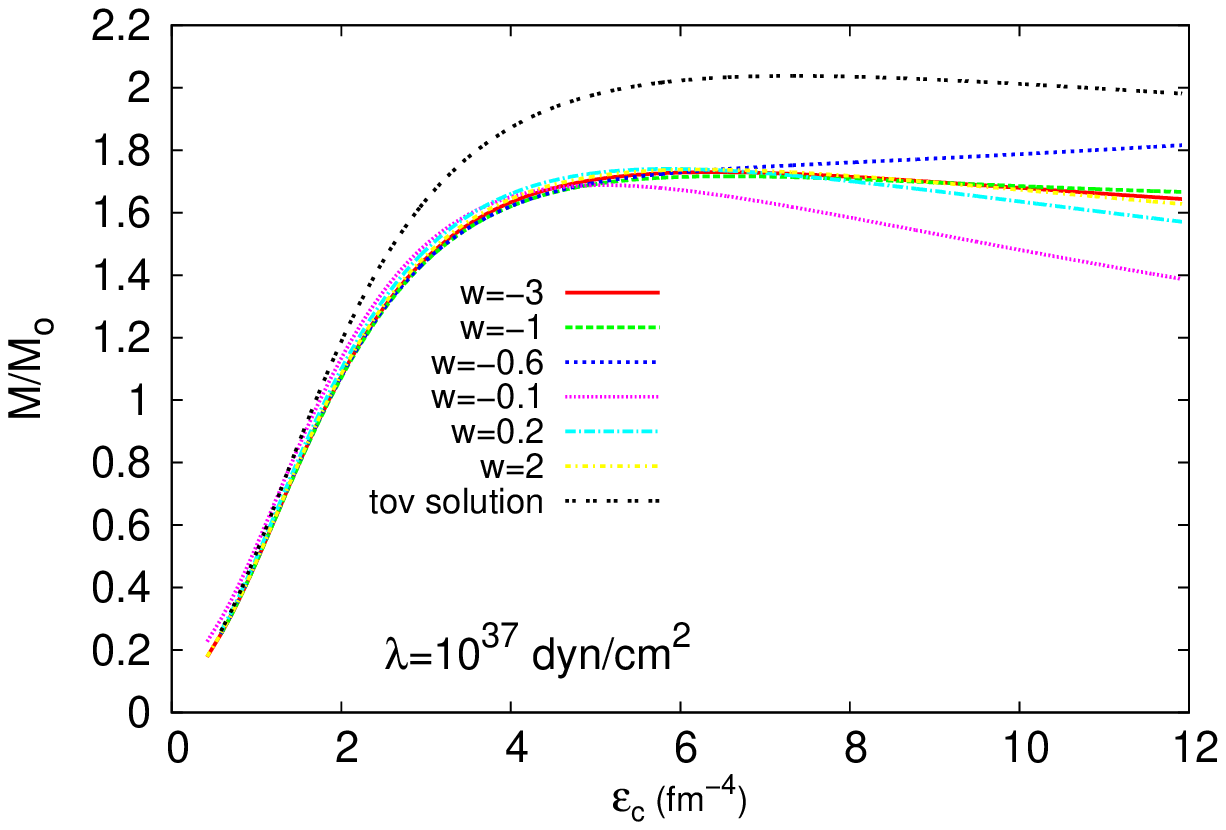}\label{fighad:c}}
\hfill
\subfloat[]{\includegraphics[width=0.45\linewidth,angle=0]{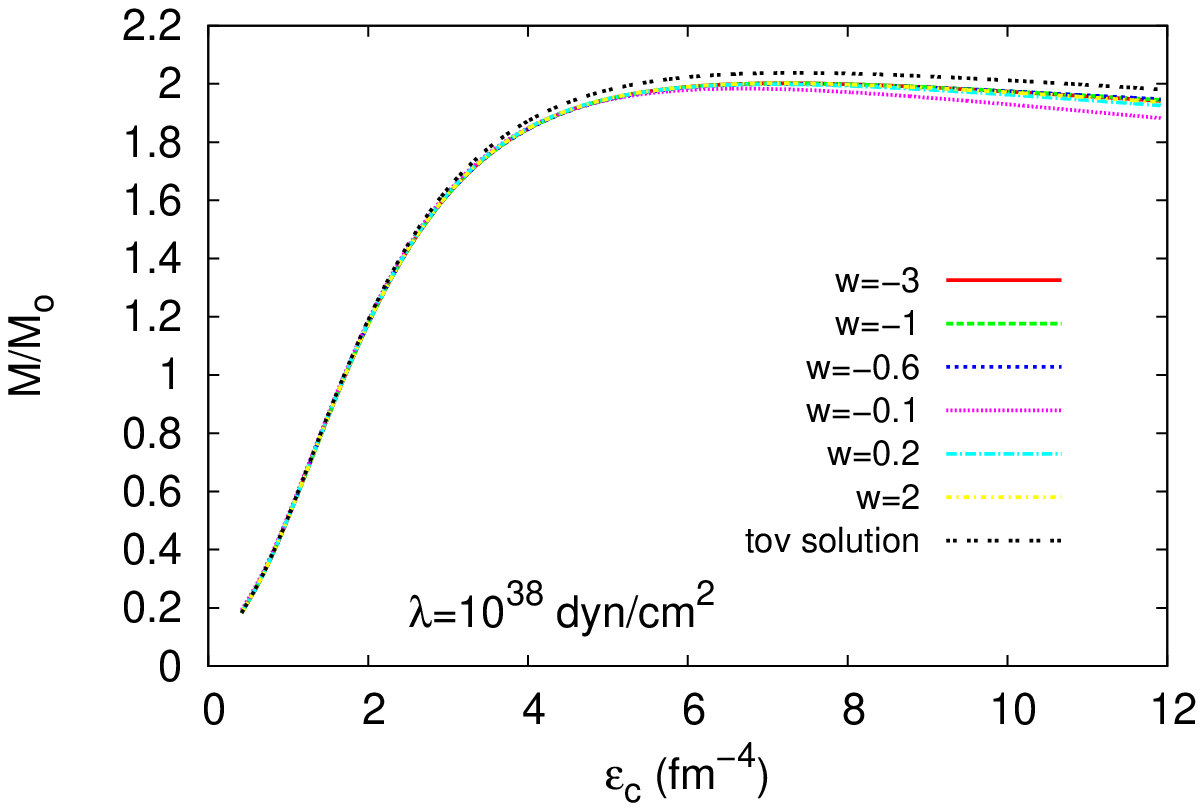}\label{fighad:d}}
\hfill
\subfloat[]{\includegraphics[width=0.45\linewidth,angle=0]{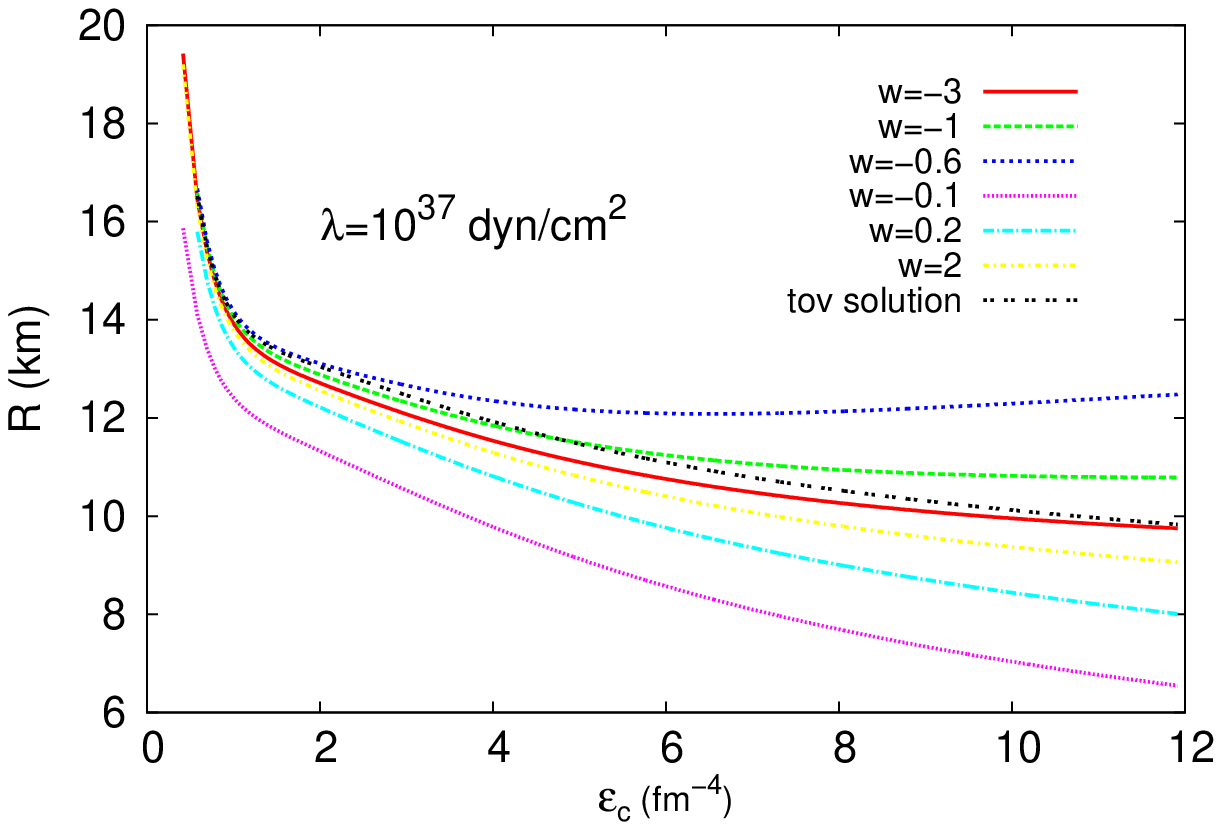}\label{fighad:e}}
\hfill
\subfloat[]{\includegraphics[width=0.45\linewidth,angle=0]{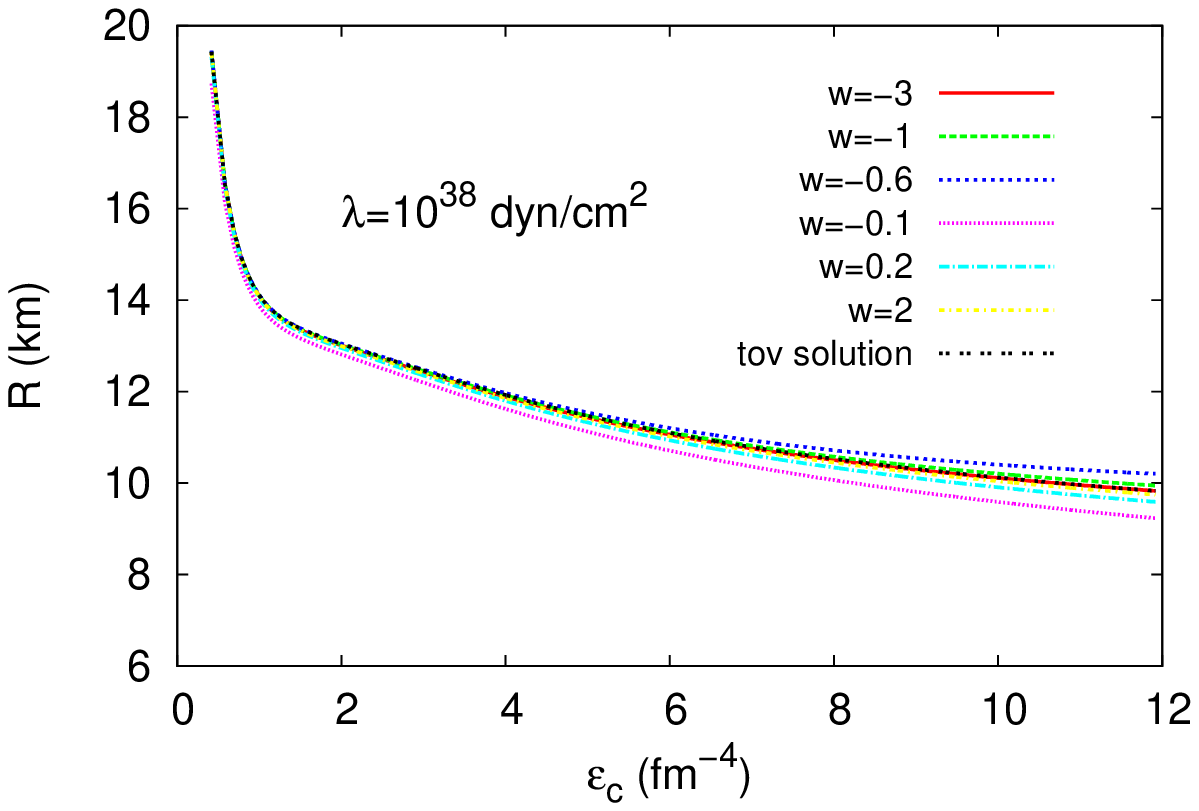}\label{fighad:f}}
\caption{Stellar properties for hadronic stars obtained with $\lambda=10^{37}$ dyn/cm$^2$ (left figures) and
$\lambda=10^{38}$ dyn/cm$^2$ (right figures) and different values of $w$ : (a), (b) mass-radius curves, (c), (d) mass in function of the central energy density and (e), (f) radius versus central energy density.}
\label{fighad}
\end{figure}

In Figure~\ref{fighad} we display the brane-TOV solutions for
hadronic stars whose equations of state include only nucleons and leptons,
necessary to enforce $\beta$-equilibrium and charge neutrality conditions.
The equation of state was obtained with the relativistic non-linear
Walecka model as in~\cite{mp2,burst}. We have fixed both the brane tension $\lambda=10^{37}$ dyn/cm$^2$
and $\lambda=10^{38}$ dyn/cm$^2$ and varied $w$. It is import to mention
that values of $w$ have been chosen in such way that the values of
radii are in accordance with some recent estimates~\cite{radii,radii2,radii3}. For a  broader range of $w$ values, all the mass-radius curves
  fall within the same interval as the ones obtained within our chosen
  range ($-3 < w < 2$) , so that with the chosen range, all possible
  mass-radius results are contemplated. Moreover, when examining equation~\eqref{eq4} for
a large value of $|w|$, one sees that the derivative of ${\cal U}$
with respect to $r$ becomes very small and hence, its contribution in
equation~\eqref{eq3} becomes negligible. If the central value of $\cal U$
  is different from zero, then
the contribution becomes a constant, but the qualitative results are equivalent.

 A first analysis of the mass-radius
curves for $\lambda=10^{37}$ dyn/cm$^2$, displayed in Figure~\ref{fighad:a} shows that the stellar masses vary in an
oscillatory way with the change of $w$ and for some values around
$w=-0.6$, the solutions may become unstable.  A necessary
condition for stability is that the derivative of the stellar mass with
respect to the central energy density must be positive \cite{glen}.

As already stated in the Introduction, the star becomes
more compact, in the sense that the masses and radii decrease for all $w$ values
as compared with the usual TOV solution. Our results of stellar properties for hadronic stars obtained with $\lambda=10^{37}$ dyn/cm$^{2}$ are summarized in Table~\ref{hadronic1}. From Figure~\ref{fighad:c}, we see that
the central energy density is practically the same for all $w$ values until the
maximum mass star is reached. Only after this point, the solutions deviate from
each other. For the sake of completeness the radii in function of the central
energy density is depicted in Figure~\ref{fighad:e}.
We then analyze the situation when $\lambda=10^{38}$ dyn/cm$^2$ shown in
Figures~\ref{fighad:b},~\ref{fighad:d} and~\ref{fighad:f}. All brane-TOV
solutions still represent slightly more compact stars than the TOV one,
but for this value of the brane tension, the stellar properties for a family
of stars are almost independent of $w$. Moreover, for this value of the brane
tension, the results are very similar
to the ones obtained with the standard TOV solution. One has to bear in mind
that these solutions are not 100\% precise due to small numerical
uncertainties. Our results of stellar properties for hadronic stars obtained with $\lambda=10^{38}$~dyn/cm$^{2}$ are summarized in Table~\ref{hadronic2}.

\begin{table}[tbp]
\centering
\begin{tabular}{|c |c c c c c c c | }
  \hline
  Hadronic  & \multicolumn{7}{ |c| }{$\lambda=10^{37}$} \\
 Stars  &  \multicolumn{7}{ |c| }{($\mathrm{dyn/cm^{2}}$)} \\
  \cline{2-8}
   & $w=-3$ & $w=-1$ & $w=-0.6$ & $w=-0.1$ & $w=-0.2$ & $w=2$ & standard TOV \\
  \hline
 $M_{\mathrm{max}}$ ($M_{0}$)  & 1.73 & 1.72 & 1.82 & 1.69 & 1.74 & 1.74 & 2.04 \\

 R (km)  & 10.64 & 11.14 & 12.48 & 9.07 & 9.81 & 10.32 & 10.70 \\

 $\varepsilon_{c}$ ($\mathrm{fm}^{-4}$)  & 6.41 & 6.56 & 11.92 & 5.10 & 5.89 & 6.25 & 7.33 \\
   \hline
\end{tabular}
\caption{\label{hadronic1} Stellar properties for hadronic stars obtained with $\lambda=10^{37}$~dyn/cm$^{2}$ and different values of $w$.}
\end{table}

\begin{table}[tbp]
\centering
\begin{tabular}{|c|ccccccc|}
  \hline
  Hadronic  & \multicolumn{7}{ |c| }{$\lambda=10^{38}$} \\
 Stars  &  \multicolumn{7}{ |c| }{($\mathrm{dyn/cm^{2}}$)} \\
  \cline{2-8}
   & $w=-3$ & $w=-1$ & $w=-0.6$ & $w=-0.1$ & $w=-0.2$ & $w=2$ & standard TOV \\
  \hline
 $M_{\mathrm{max}}$ ($M_{0}$)  & 2.00 & 2.00 & 2.00 & 1.98 & 2.00 & 2.00 & 2.04 \\

 R (km)  & 10.70 & 10.75 & 10.88 & 10.46 & 10.60 & 10.67 & 10.70 \\

 $\varepsilon_{c}$ ($\mathrm{fm}^{-4}$)  & 7.21 & 7.24 & 7.24 & 6.70 & 7.04 & 7.16 & 7.33 \\
   \hline
\end{tabular}
\caption{\label{hadronic2} Stellar properties for hadronic stars obtained with $\lambda=10^{38}$~dyn/cm$^{2}$ and different values of $w$.}
\end{table}

\begin{figure}[tbp] 
\centering
\includegraphics[width=0.55\linewidth,angle=0]{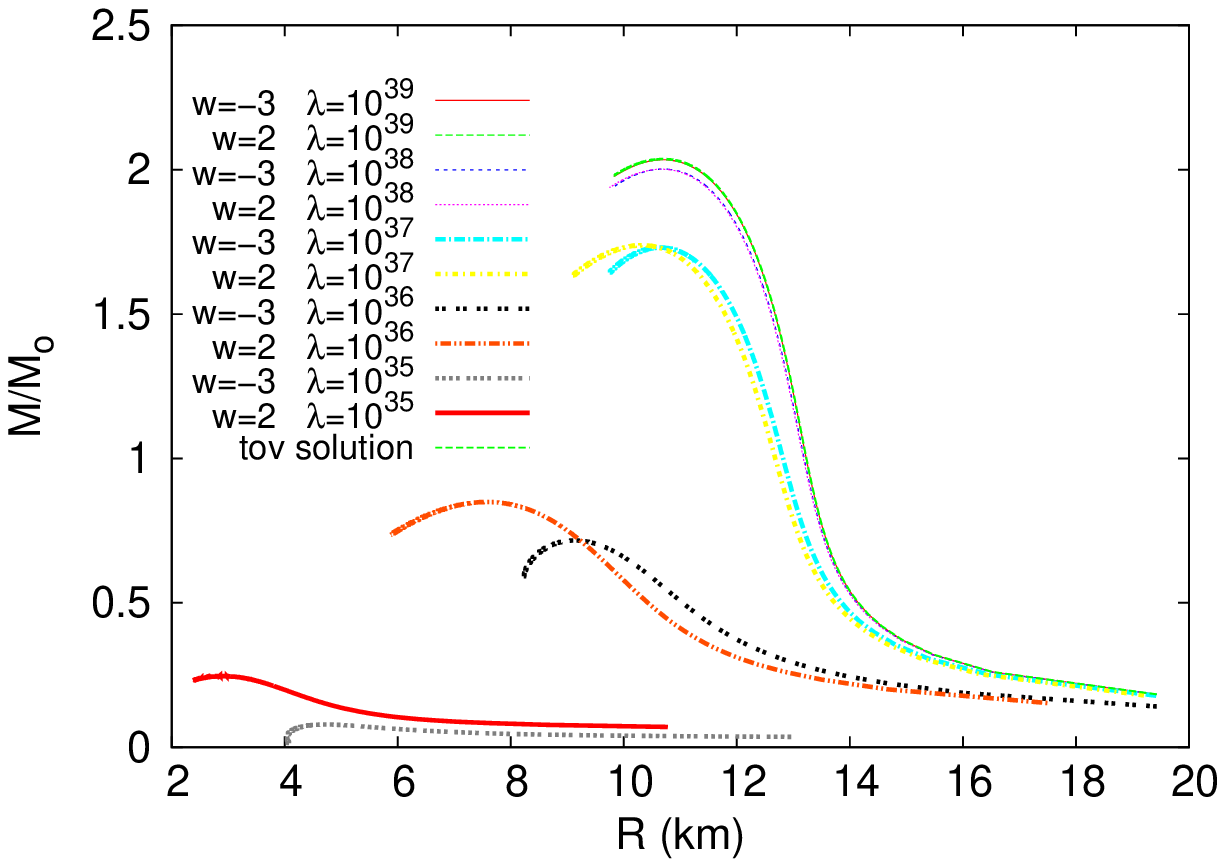}\\
\caption{\label{fighad2} Mass radius curves obtained from the same relativistic equation of
state as in figure~\ref{fighad} for different values of $\lambda$ and
two values of $w$.}
\end{figure}

\begin{figure}[tbp]
\centering
\includegraphics[width=0.55\linewidth,angle=0]{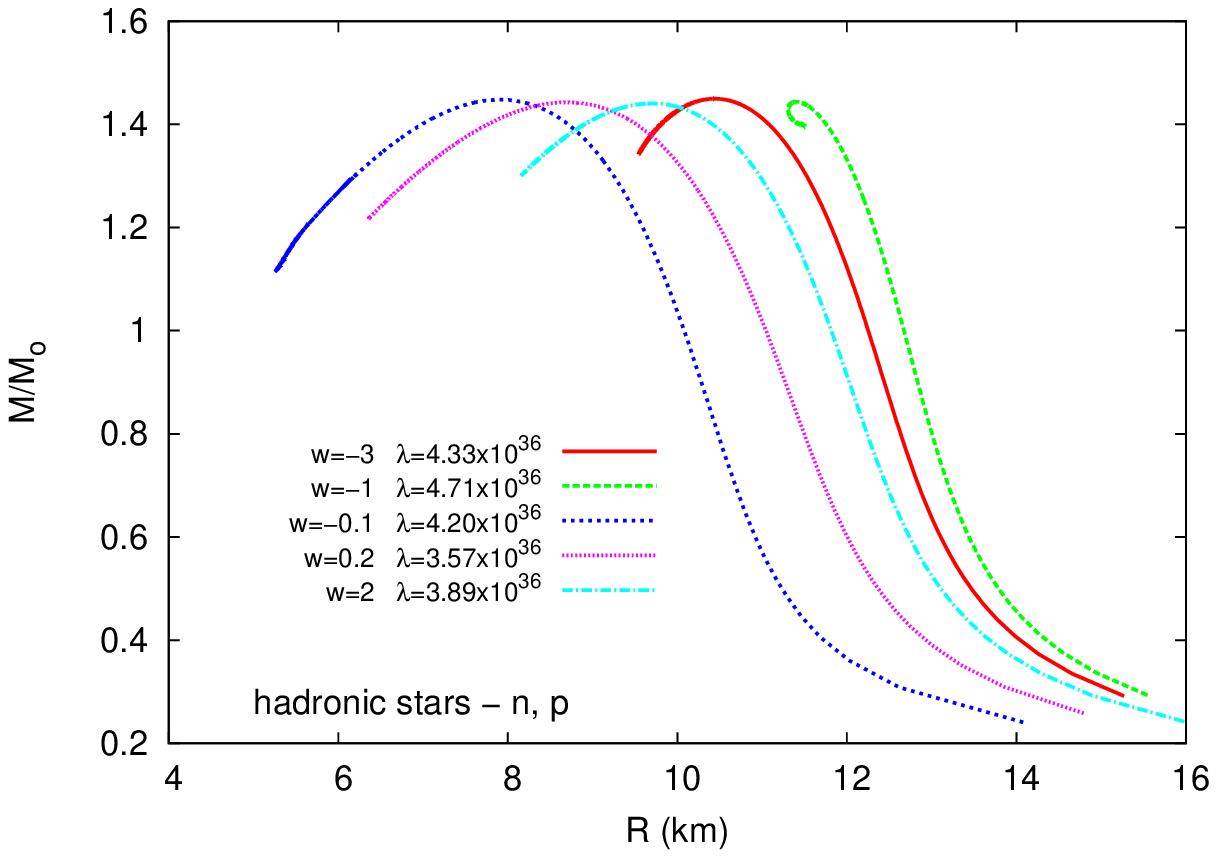}\\
\caption{\label{fighad3} Mass radius curves with maximum mass $1.44~M_{\odot}$ obtained
 from the same relativistic equations of state as in figure~\ref{fighad}.}
\end{figure}

We then vary $\lambda$ from $10^{35}$ to $10^{39}$ dyn/cm$^2$ for two fixed
values of $w$, one negative ($w=-3$) and one positive ($w=2$). The mass-radius
curves are shown in Figure~\ref{fighad2}, from
where we see that the stellar masses increase with the increase of $\lambda$
and tend to converge to their maximum values around
$\lambda=10^{38}$ dyn/cm$^2$.
On the other hand, when this value is achieved, the results
depend only slightly on the $w$ values, as already seen in
Figures~\ref{fighad:b},~\ref{fighad:d} and~\ref{fighad:f}.
An interesting aspect related to these
solutions is the fact that $\lambda$ clearly controls the values of the
maximum star masses, while $w$ influences the corresponding radii.
For small values of the brane tension, the brane-TOV solutions become either unstable or produce very low maximum masses, what is not expected from
astronomical observations. Thus, we can obtain both a lower limit
for the brane tension and a value for which the usual TOV
results are obtained, as already discussed in~\cite{germani,garcia}, in a simplified
context, where a perfect fluid was used instead of a more realistic equation of
state. In the present work, we establish a range for $\lambda$ in between
$3.89 \times 10^{36} < \lambda < 10^{38}$ dyn/cm$^2$, the lower limit is obtained in such a way that
at least a $1.44~ M_{\odot}$ star can be achieved. Of course, as discussed above
and seen in Figure~\ref{fighad3}, these tensions depend on the $w$ values.

\begin{figure}[tbp]
\centering
\subfloat[]{\includegraphics[width=0.45\linewidth,angle=0]{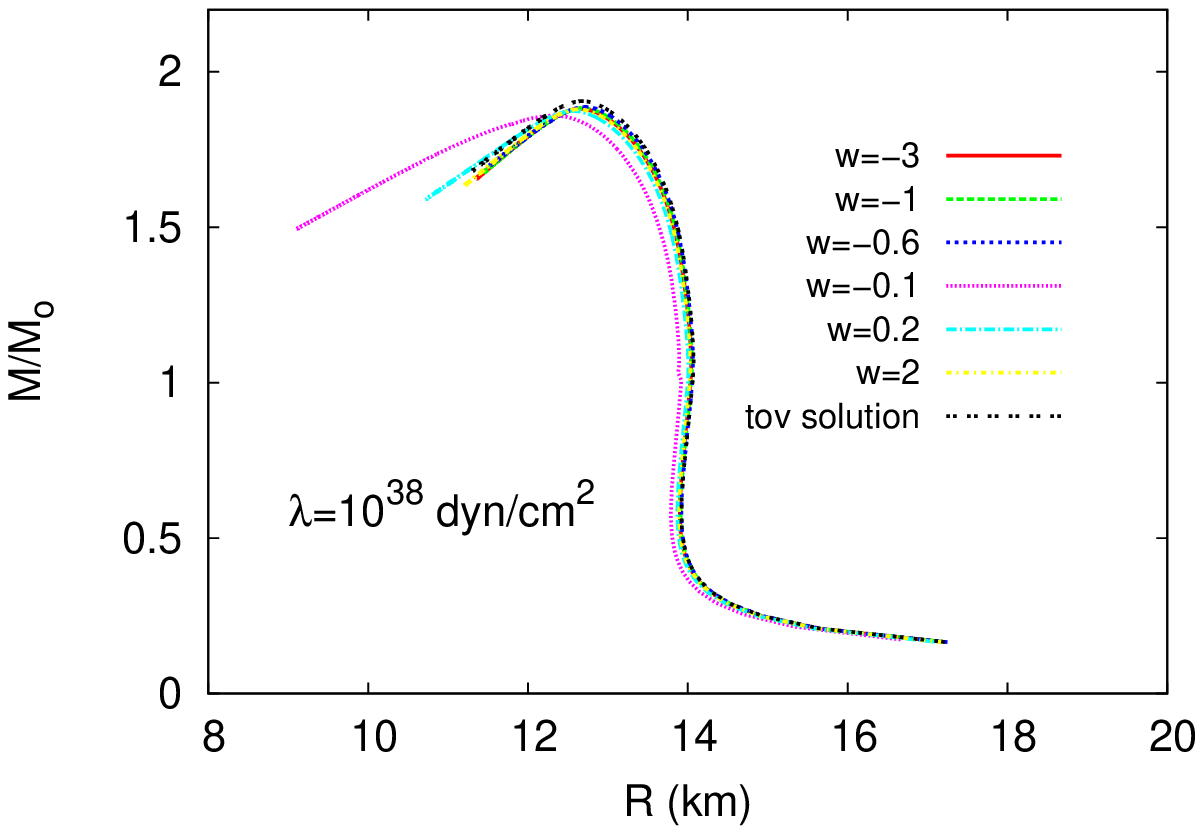}\label{fighyb:a}}
\hfill
\subfloat[]{\includegraphics[width=0.45\linewidth,angle=0]{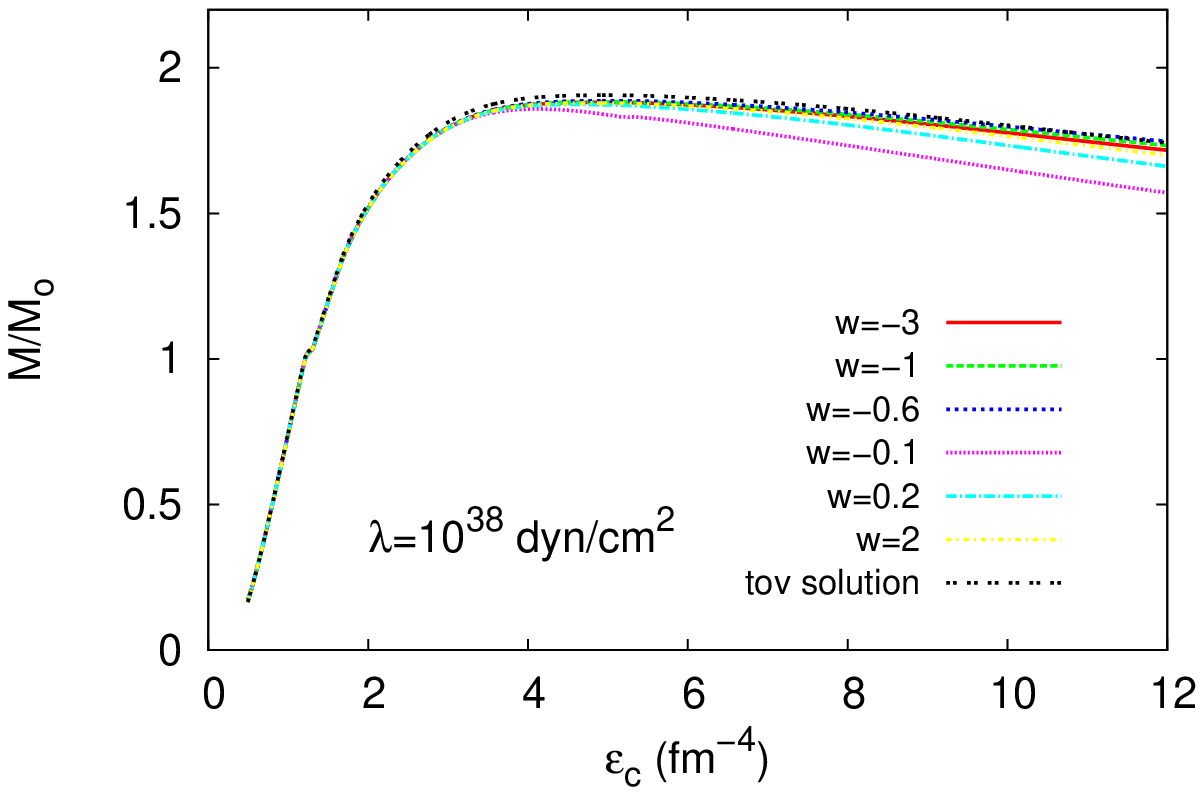}\label{fighyb:b}}
\hfill
\subfloat[]{\includegraphics[width=0.45\linewidth,angle=0]{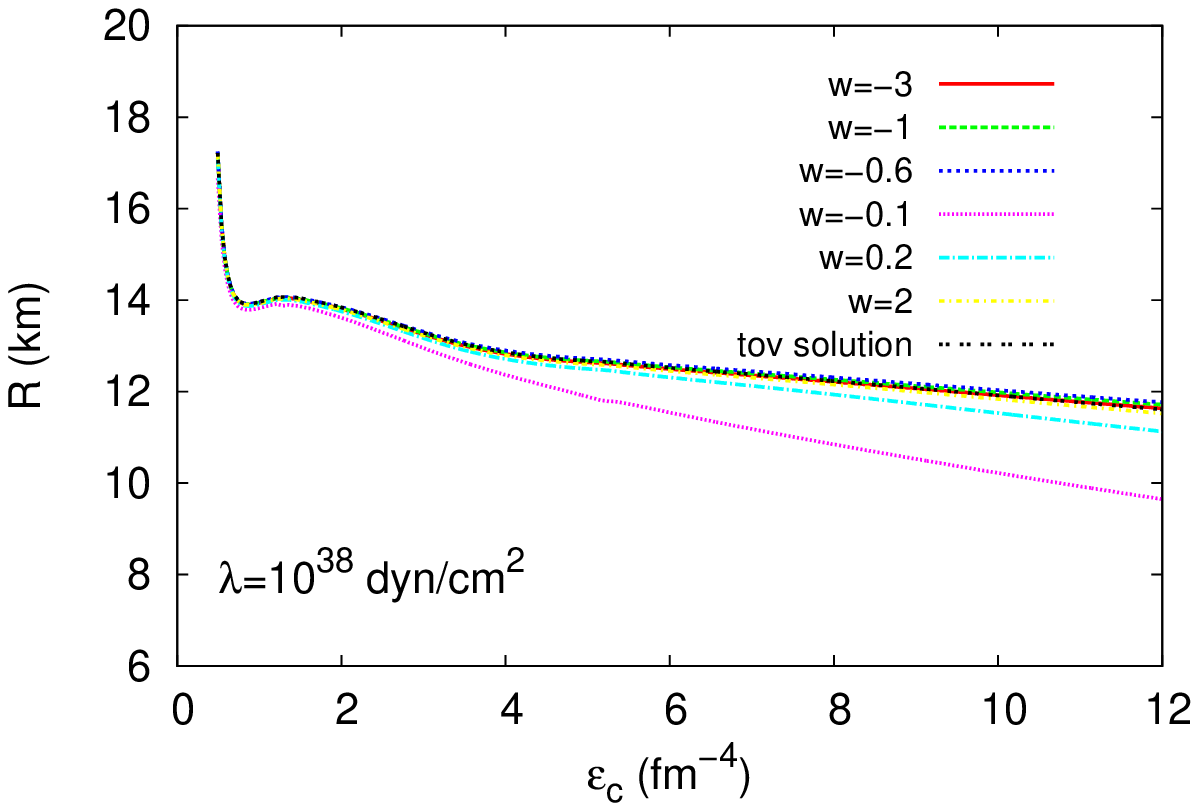}\label{fighyb:c}}
\caption{Hybrid star properties obtained with
$\lambda=10^{38}$ dyn/cm$^2$ and different values of $w$ :
(a) mass-radius curves, (b) mass in function of the central energy density
and (c) radius versus central energy density.}
\label{fighyb}
\end{figure}

Next we move to the study of hybrid stars. The equation of state we use as
input to the brane-TOV equations is obtained from the non-linear
Walecka model for the hadron phase, the Nambu-Jona-Lasinio model for the quark
phase and a Gibbs construction for the mixed phase as in~\cite{mp1,mp1_1}. In face
of the results obtained for hadronic stars, we have chosen
$\lambda=10^{38}$ dyn/cm$^2$. Our results of stellar properties for hybrid stars are summarized in Table~\ref{hybrid1}. From the Figure~\ref{fighyb}, we can see that the general qualitative behavior is the same as for hadronic stars, except for the kinks in all curves related to the
transitions from one phase to the other inside the star structure.
However, for certain values of $w$ ($w=-0.1$, for instance), there are still
sensitive deviations from the standard TOV results, which means that,
even for the limit brane tension, the results still depend on the value of $w$
if hybrid stars are considered.

\begin{table}[tbp]
\centering
\begin{tabular}{|c|ccccccc|}
  \hline
  Hybrid  & \multicolumn{7}{ |c| }{$\lambda=10^{38}$} \\
 Stars  &  \multicolumn{7}{ |c| }{($\mathrm{dyn/cm^{2}}$)} \\
  \cline{2-8}
   & $w=-3$ & $w=-1$ & $w=-0.6$ & $w=-0.1$ & $w=-0.2$ & $w=2$ & standard TOV \\
  \hline
 $M_{\mathrm{max}}$ ($M_{0}$)  & 1.88 & 1.88 & 1.89 & 1.86 & 1.87 & 1.88 & 1.91 \\

 R (km)  & 12.65 & 12.68 & 12.72 & 12.31 & 12.56 & 12.62 & 12.67 \\

 $\varepsilon_{c}$ ($\mathrm{fm}^{-4}$)  & 4.86 & 4.95 & 5.00 & 4.11 & 4.58 & 4.84 & 4.97 \\
   \hline
\end{tabular}
\caption{\label{hybrid1} Stellar properties for hybrid stars obtained with $\lambda=10^{38}$~dyn/cm$^{2}$ and different values of $w$.}
\end{table}

\begin{table}[tbp]
\centering
\begin{tabular}{|c|ccccccc|}
  \hline
  Quark  & \multicolumn{7}{ |c| }{$\lambda=10^{38}$} \\
 Stars  &  \multicolumn{7}{ |c| }{($\mathrm{dyn/cm^{2}}$)} \\
  \cline{2-8}
   & $w=-3$ & $w=-1$ & $w=-0.6$ & $w=-0.1$ & $w=-0.2$ & $w=2$ & standard TOV \\
  \hline
 $M_{\mathrm{max}}$ ($M_{0}$)  & 2.29 & 2.29 & 2.30 & 2.24 & 2.28 & 2.29 & 2.32 \\

 R (km) & 11.96 & 11.97 & 12.01 & 11.77 & 11.90 & 11.94 & 11.98 \\

 $\varepsilon_{c}$ ($\mathrm{fm}^{-4}$)  & 4.50 & 4.56 & 4.56 & 4.12 & 4.41 & 4.49 & 4.60 \\
   \hline
\end{tabular}
\caption{\label{quark1} Stellar properties for quark stars obtained with $\lambda=10^{38}$~dyn/cm$^{2}$ and different values of $w$.}
\end{table}

\begin{figure}[tbp]
\centering
\subfloat[]{\includegraphics[width=0.45\linewidth,angle=0]{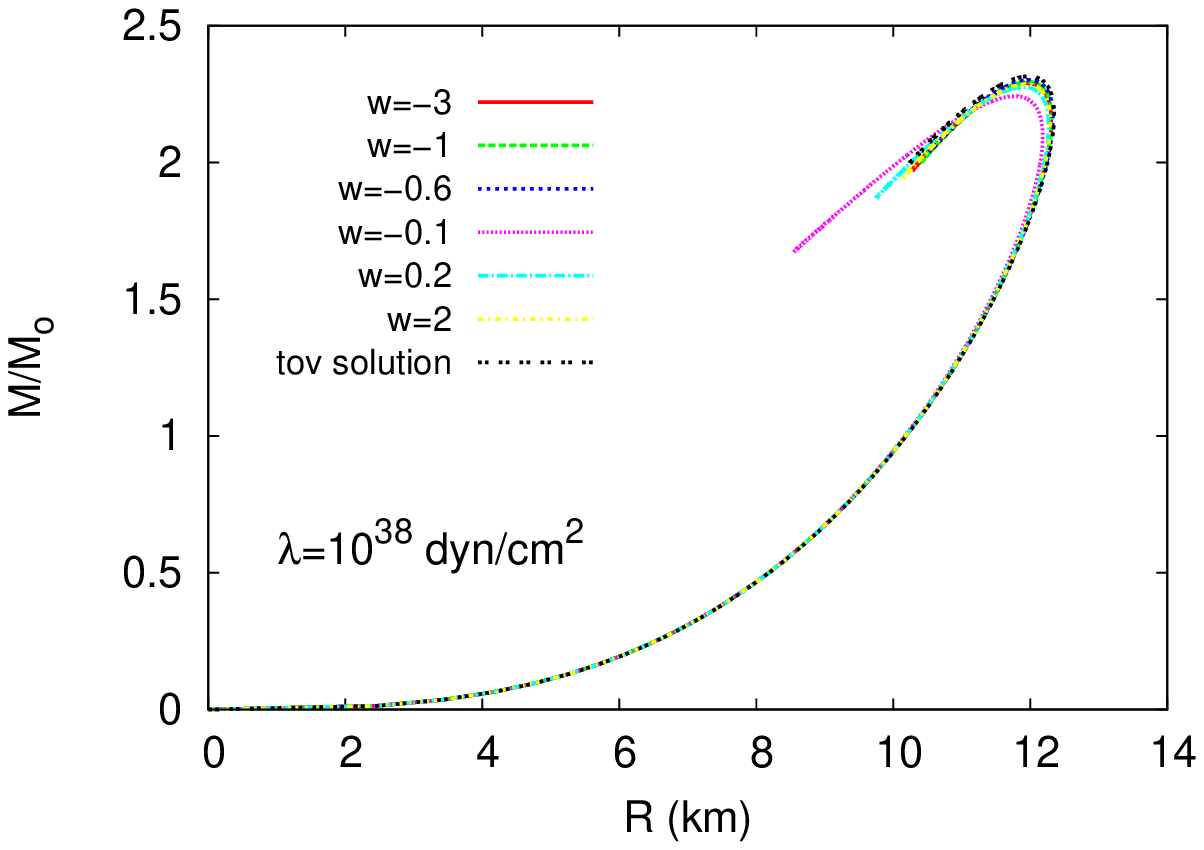}\label{figquark:a}}
\hfill
\subfloat[]{\includegraphics[width=0.45\linewidth,angle=0]{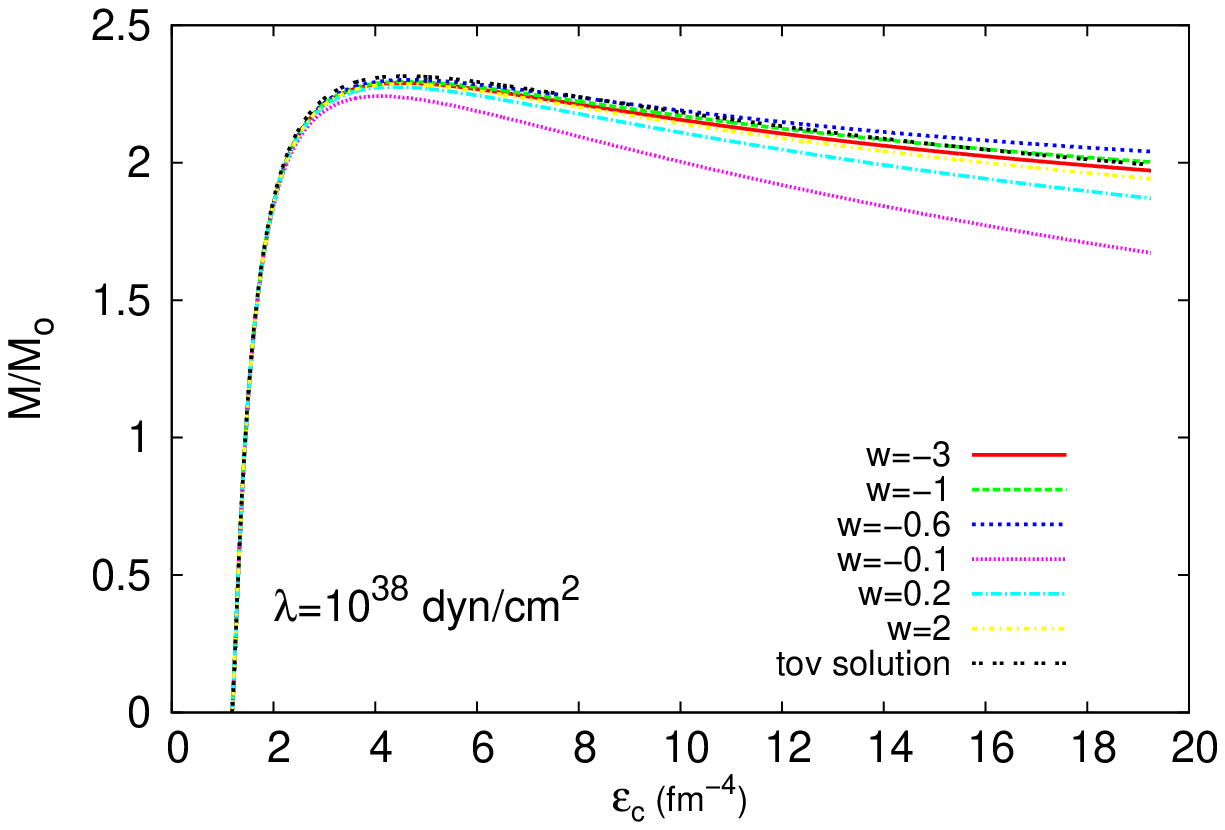}\label{figquark:b}}
\hfill
\subfloat[]{\includegraphics[width=0.45\linewidth,angle=0]{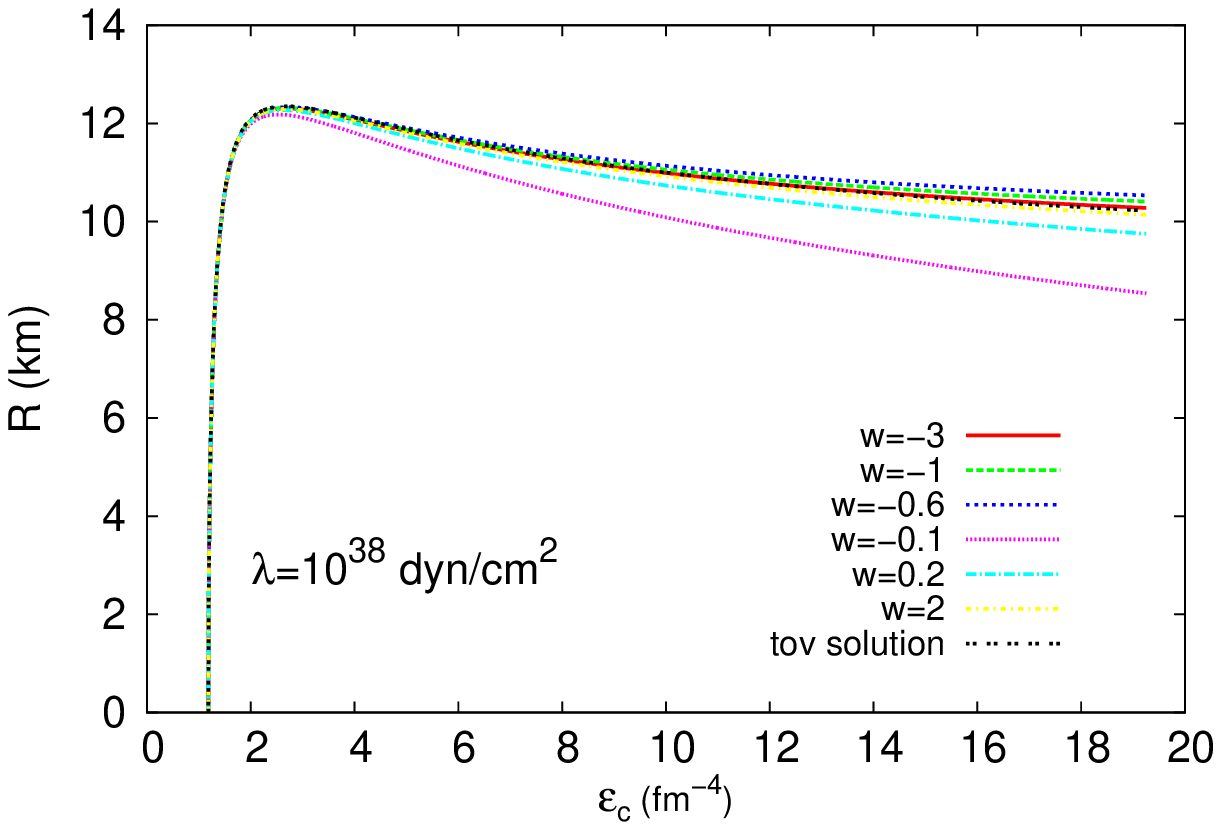}\label{figquark:c}}
\caption{Quark star properties obtained with
$\lambda=10^{38}$ dyn/cm$^{2}$ and different values of $w$ :
(a) mass-radius curves, (b) mass in function of the central energy density
and (c) radius versus central energy density.}
\label{figquark}
\end{figure}

To finish our analysis, we look at an equation of state that is used to
describe quark stars. A model for quark matter that yields maximum masses
of the order of $2~M_{\odot}$ is the quark mass density dependent model~\cite{qmdd,qmdd_2} and we use an equation of state taken from~\cite{james}.
Once again we see, from Figure~\ref{figquark},
that for the chosen value of $\lambda=10^{38}$~dyn/cm$^2$,
the TOV solutions are practically reproduced for a certain range of $w$ values,
as in the case of hybrid stars. Our results of stellar properties for quark stars are summarized in Table~\ref{quark1}. A point worth mentioning is the behavior of the
quark star radius as a function of the energy density shown in
Figure~\ref{figquark:c}, very similar to the mass-energy density, due to the fact
that quark stars present a finite density at their surface contrary to hadronic
and hybrid stars, which have zero densities at zero pressure surface points.
These different physical structures are seen when one compares
Figures~\ref{fighad:a} and~\ref{fighyb:a} with Figure~\ref{figquark:a}.

\section{Final conclusions}
\label{sec5}

In the present work we have revisited the brane-TOV solutions
discussed in~\cite{germani,garcia} for realistic equations of state
normally used in the literature and compared the results with the ones
obtained from the standard TOV solutions. In order to solve
satisfactorily the brane-TOV equations we need an equation of state
$p=p(\epsilon)$ and additionally an equation of state-like relation
$\mathcal{P}=\mathcal{P}(\mathcal{U})$. We have assumed that the Weyl
terms obey the simplest relation $\mathcal{P}=w\mathcal{U}$.
 We have then chosen appropriate EOS for the description of
hadronic, hybrid and quark stars as input to the brane-TOV equations.
An interesting aspect related to our results is the fact that the
brane tension $\lambda$ clearly controls the values of the maximum
star masses, while $w$ influences the corresponding radii. We have
established a range for $\lambda$ in between
$3.89\times10^{36}<\lambda<10^{38}$~dyn/cm$^{2}$. The lower limit is
obtained in such a way that at least a $1.44~M_\odot$ star can be
achieved. On the other hand, there is a value for which
the solutions encountered reproduce the standard TOV solutions. This fact means that, as far as the equations of state survive the observational constraints when the macroscopic properties are computed from the usual TOV equations, they are also suitable as input to the brane-TOV equations. Once the maximum brane tension value is attained, the results are practically independent of the value of $w$ for hadronic stars and very little dependent for hybrid and quarks stars.

It is very important to make some comments on the possible values of neutron stars radii. Based on chiral effective theory, the authors of Ref.~\cite{radii} estimate the radii of the canonical $1.4\,M_\odot$ neutron star to lie in the range 9.7$-$13.9 km. More recently, two different analysis of five quiescent low-mass X-ray binaries in globular clusters resulted in different ranges for neutron star radii.
The first one, in which it was assumed that all neutron stars have the same radii, predicted that they should lie in the range $R=9.1^{+1.3}_{-1.5}$~\cite{radii2}. The second calculation, based on a Bayesian analysis, foresees radii of all neutron stars to lie in
between 10 and 13.1 km~\cite{radii3}. If one believes those are
definite constraints, all hadronic, hybrid and quark stars with the
choice of EOS studied in the present work survive the observational
constraints for values of $w$ in between $-3<w<2$ (excluding an
interval of $w$ values around -0.6) for $\lambda=10^{37}$~dyn/cm$^{2}$).
For other values of $w$, the brane-TOV solutions produce mass-radius results which
fall within the same interval as the ones obtained within our chosen range.
We can conclude that one advantage of using the brane-TOV equations is that the radii can be adjusted to smaller values, as seen, for instance in Figure~\ref{fighad:a}.

\acknowledgments

L. B. Castro would like to thank Professor Dr. J. M. Hoff da Silva for useful discussions. The authors are indebted to the anonymous referee for an excellent and constructive review. This work was partially done during a visit (L. B. Castro) to UNESP-Campus de Guaratinguet\'{a}. This work was supported by CNPq (Brazil), Capes (Brazil) and
FAPESC (Brazil) under project 2716/2012, TR 2012000344.



\end{document}